\documentclass[submission,copyright,creativecommons]{eptcs}
\providecommand{\event}{FTSCS 2012} % Name of the event you are submitting to
\usepackage{graphicx}
\usepackage{subfig}
\DeclareCaptionType{copyrightbox}
\title{Formal Verification, Engineering and Business Value\\
}
\author{Ralf Huuck
\institute{NICTA\\ Sydney, Australia}
\institute{School of Computer Science and Engineering\\
University of New South Wales\\
Sydney, Australia}
\email{ralf.huuck@nicta.com.au}
}
\def\titlerunning{Formal Verification, Engineering and Business Value}
\def\authorrunning{R.Huuck}
\begin{document}
\maketitle

\begin{abstract}
How to apply automated verification technology such as model checking 
and static program analysis to millions of lines of embedded C/C++ code? 
How to package this technology in a way that it can be used by software 
developers and engineers, who might have no background in formal verification? 
And how to convince business managers to actually pay for such a software? 
This work addresses a number of those questions. Based on our own experience on 
developing and distributing the Goanna source code analyzer for detecting software 
bugs and security vulnerabilities in C/C++ code, we explain the underlying technology
 of model checking, static analysis and SMT solving, steps involved in creating industrial-proof 
tools.
\end{abstract}

\section{Motivation}

Formal verification has come a long way from being a niche domain for mathematicians and 
logicians to an accepted practice, at least in academia, and frequently being taught in undergraduate 
courses. Moreover, starting out from a pen-and-paper approach, a range of supporting software 
tools have been developed over time including specification tools for (semi-)formal languages such as UML,
Z or various process algebrae, interactive theorem-provers for formal specification, proof-generation and 
verification, as well as a large number of algorithmic software tools for model checking, run-time verification, static analysis and SMT solving to name a few  \cite{dkw2008}. 

Despite all the effort, however, there has been only limited penetration of verification tools
into industrial environments, mostly remaining confined to the respective R\&D laboratories of 
larger corporations, defense projects or selective avionics work. The use of verification tools by the average software engineer is rare and typically stops at formal techniques built into the compiler or debugger.

One of the key motivations for our work has been to contribute to some of the technology transfer from classical academic domains to industrial applications and use. In particular, we have been working on bringing verification techniques such as model checking \cite{queillesifakis82,clarkeemerson81}, abstract interpretation \cite{CousotCousot79-1} and SMT solving \cite{DeMoura:2011ff} to professional software engineers. 

To make verification technology applicable, we believe a number of core principles must be met: First of all, any verification tool has to be so simple to use that it does not require much or any learning from users outside the formal methods domain. Secondly, the performance of the tool has to match the typical workflow of the end-user. This means, if the end-user is accustomed to doing things in a particular order, those steps should remain largely the same. Moreover, run-time performance of any additional analysis or verification should be similar to existing processes. Finally, and most importantly, a new analysis tool should provide real value to an end-user. This means, it should deliver information or a degree of reliability that was previously not available, making the use of the tool worthwhile.

\section{Software Tool Challenges}

The result of our endeavor is \emph{Goanna} \cite{FHJLR07}, an automated software analysis tool for detecting software bugs, code anomalies and security vulnerabilities in C/C++ source code. Goanna is designed to be run at compile-time and does not require any annotations, code modifications or user interaction. Moreover, the tool can directly be integrated into common development environments such as Eclipse, Visual Studio or build systems based on, e.g., Makefiles. To achieve acceptance in industry, all formal techniques are hidden behind a typical programmer's interface, all of C/C++ is accepted as input (even, e.g., Microsoft specific compiler extensions) and scalability had to be ensured for many millions of lines of code in reasonable time.

To achieve this, a number of trade-offs had to be made: While using a range of formal verification techniques, Goanna is not a verification tool as such, but rather a bug detection tool. This means, it does no conclusively show the absence of errors, but does its best to find certain classes of bugs. Moreover, Goanna comes by default with a fixed set of pre-defined checks for errors such as buffer overruns, memory leaks, NULL-pointer dereferences, arithmetic errors, or C++ copy control mistakes as well as with support for certain safety-critical coding standards such as MISRA \cite{motor1998guidelines} or CERT \cite{seacord2008cert} totaling over 200 individual checks. To achieve reasonable performance that is of the same order as the compiler, a number of assumptions and approximations (as well as refinements) are made. Naturally, this leads to missed errors (false negatives) as well as to spurious errors (false positives). Finding the right balance between precision, speed and number of supported checks is very much an engineering art supported by new verification and abstraction techniques. This work will focus on some key insights:

\begin{description}
\item[Core Analysis] The core of our program analysis is based on CTL model checking. In particular, most static analysis tasks such as NULL-pointer detection, buffer overrun analysis and memory leak detection are outsourced to an explicit state CTL model checker and an abstract interpretation framework. The approach of model checking the distribution of syntactic elements in C/C++ code creates particular requirements such as handling many small verification tasks. We implemented our own explicit state model checker outperforming some existing state-of-the-art implementation by up to an order of magnitude.

\item[Refinements \& Heuristics] Goanna, in its academic version, integrates with SMT solvers and applies a refinement loop for distinguishing spurious errors from real ones \cite{Junker:2012}. Moreover, a substantial engineering effort has been done to apply heuristics matching common programmer patterns and expectations to keep the overall false-alarm rate to a minimum.

\item[Engineering Challenges] While formal methods tend to be precise and define right and wrong, software practice is often much less rigorous and one person's bug is another person's feature. Moreover, C/C++ are complex languages with often tricky semantics and, more importantly, are neither stable with new features being added through the C++11 standard or compiler extensions, nor are these languages the same for everyone, but rather determined by how the individual compiler manufacturer supports and interprets C/C++.

Internally, we have to deal with huge number of verification tasks. This can range to hundreds of millions of model checking specifications for larger projects and sometimes billions of pre-processing pattern matching queries. This is typically contrary to an academic environment, where one is often interested in a few complex verification tasks only. Apart from new algorithms this requires dedicated engineering solutions for efficient caching, incremental analysis, inter-function summaries and databases for storage and look-up between runs. 

\item[Business Value] Interestingly, software analysis plays only a small role in the overall software development lifecycle and general product development process. Most importantly, organizations need to get products to market as fast as possible with as many features as possible. Bugs are of concern if they lead to critical failure and therefore a loss in business. A bug that never or very rarely manifests is often not of primary concern. However, of importance is being able to track the overall software quality and observe trending, see levels of compliance to standards and coding guidelines, and to obtain reports that can be understood by management. Most importantly, it is worth to note that the people making purchasing decisions are typically different from the software developers and motivated by quite different reasons.
\end{description}

\begin{figure}[t]
\center
\includegraphics[width=0.4\textwidth]{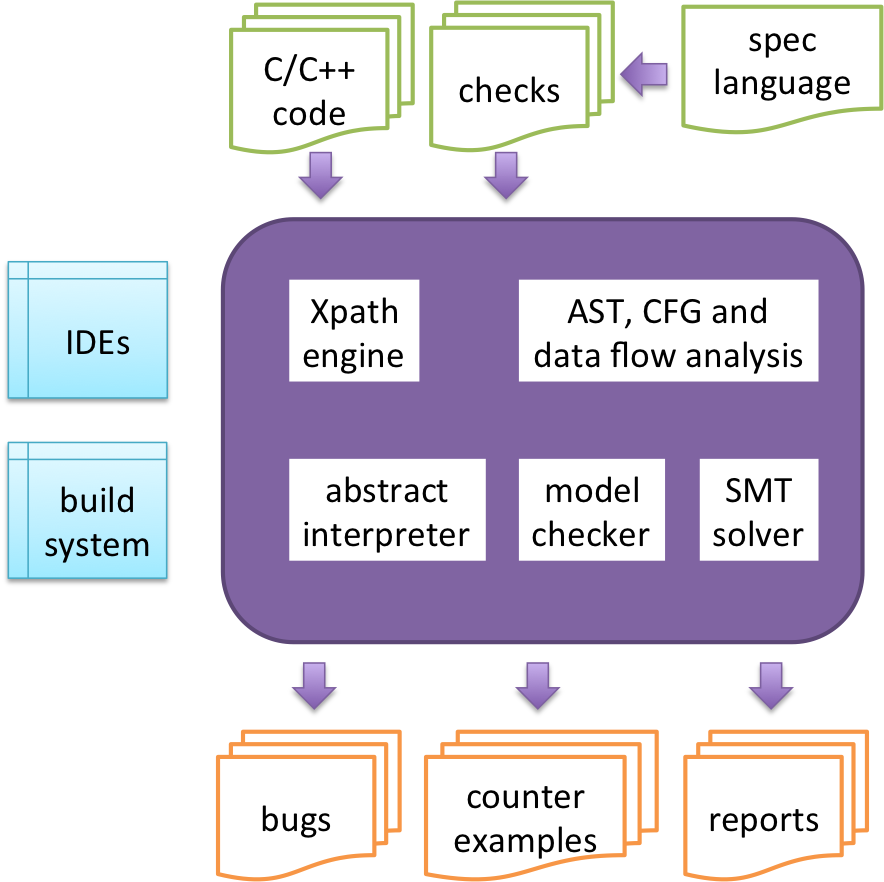}
\caption{High-level Goanna Architecture}
\label{fig:arch}
\end{figure}

Figure~\ref{fig:arch} depicts the high-level architecture of our software analysis tool. At the core are the different analysis engines for abstract interpretation, model checking, SMT solving and pattern matching. As input the tool takes C/C++ projects and a set of specifications that are written in our own domain-specific language and pre-defined for standard users. Goanna can be run on the command line, integrates with IDEs and can be embedded in the build system. It outputs warning messages classified by severity, displays error traces based on counter-examples, and can create summary reports.

\section{Lessons}

While it is a difficult endeavor to target industry with formal methods there are a range of valuable lessons that can be learned. This starts out with understanding the problem domain, the actual demands and challenges faced by the end-user as well as understanding their processes and potential \emph{value add} by tools. Technology is only one piece of a larger puzzle.

Moreover, software developers face time constraints and pressure to deliver results. Learning a new tool from scratch is not only a process change management shies away from, but typically comes with some kind of formal training in many organization often being much more costly than the actual software tool. Hence, any tool that integrates into existing IDEs and processes that work ``with the click of a button'' gains much easier acceptance.

From a formal methods researcher's point of view demands are often sobering. Real life applications require to deal with huge amounts of code in relatively short time. This will require not only abstractions, but all various heuristics, both on an algorithmic level as well as on an engineering level predicting \emph{typical} developer behavior. Generally, the engineering requirements can easily become a roadblock for the acceptance of more advanced formal analysis techniques.

Finally, with much less effort we believe that many existing academic tools could see a larger industry uptake  by following sometimes  simple rules. This includes: providing a clear and end-user friendly license agreement, providing a well documented introduction to a tool, as well as a delivering degree of reliability that the software works in \emph{most} cases. This is, however, easier said than done as academic environments typically do no provide for dedicated software engineers to develop and maintain tools over longer period of time. 

\bibliographystyle{eptcs}
\bibliography{biblio}

\begin{thebibliography}{1}
\providecommand{\bibitemdeclare}[2]{}
\providecommand{\surnamestart}{}
\providecommand{\surnameend}{}
\providecommand{\urlprefix}{Available at }
\providecommand{\url}[1]{\texttt{#1}}
\providecommand{\href}[2]{\texttt{#2}}
\providecommand{\urlalt}[2]{\href{#1}{#2}}
\providecommand{\doi}[1]{doi:\urlalt{http://dx.doi.org/#1}{#1}}
\providecommand{\bibinfo}[2]{#2}

\bibitemdeclare{misc}{motor1998guidelines}
\bibitem{motor1998guidelines}
\bibinfo{author}{Motor Industry Software~Reliability \surnamestart
  Association\surnameend} et~al. (\bibinfo{year}{1998}):
  \emph{\bibinfo{title}{Guidelines for the use of the C language in vehicle
  based software}}.

\bibitemdeclare{inproceedings}{clarkeemerson81}
\bibitem{clarkeemerson81}
\bibinfo{author}{Edmund~M. \surnamestart Clarke\surnameend} \&
  \bibinfo{author}{E.~Allen \surnamestart Emerson\surnameend}
  (\bibinfo{year}{1982}): \emph{\bibinfo{title}{Design and Synthesis of
  synchronization skeletons for branching time temporal logic}}.
\newblock In: {\sl \bibinfo{booktitle}{Logics of Programs Workshop, New York,
  May 1981}}, {\sl \bibinfo{series}{LNCS}} \bibinfo{volume}{131},
  \bibinfo{publisher}{Springer Verlag}, pp. \bibinfo{pages}{52--71},
  \doi{10.1007/BFb0025774}.

\bibitemdeclare{inproceedings}{CousotCousot79-1}
\bibitem{CousotCousot79-1}
\bibinfo{author}{P{.} \surnamestart Cousot\surnameend} \& \bibinfo{author}{R{.}
  \surnamestart Cousot\surnameend} (\bibinfo{year}{1979}):
  \emph{\bibinfo{title}{Systematic design of program analysis frameworks}}.
\newblock In: {\sl \bibinfo{booktitle}{Conference Record of the Sixth Annual
  ACM SIGPLAN-SIGACT Symposium on Principles of Programming Languages}},
  \bibinfo{publisher}{ACM Press, New York, NY}, \bibinfo{address}{San Antonio,
  Texas}, pp. \bibinfo{pages}{269--282}, \doi{10.1145/567752.567778}.

\bibitemdeclare{article}{DeMoura:2011ff}
\bibitem{DeMoura:2011ff}
\bibinfo{author}{Leonardo \surnamestart De~Moura\surnameend} \&
  \bibinfo{author}{Nikolaj \surnamestart Bj{\o}rner\surnameend}
  (\bibinfo{year}{2011}): \emph{\bibinfo{title}{{Satisfiability modulo
  theories: introduction and applications}}}.
\newblock {\sl \bibinfo{journal}{Communications of the ACM}}
  \bibinfo{volume}{54}(\bibinfo{number}{9}), \doi{10.1145/1995376.1995394}.

\bibitemdeclare{article}{dkw2008}
\bibitem{dkw2008}
\bibinfo{author}{Vijay \surnamestart D'Silva\surnameend},
  \bibinfo{author}{Daniel \surnamestart Kroening\surnameend} \&
  \bibinfo{author}{Georg \surnamestart Weissenbacher\surnameend}
  (\bibinfo{year}{2008}): \emph{\bibinfo{title}{A Survey of Automated
  Techniques for Formal Software Verification}}.
\newblock {\sl \bibinfo{journal}{IEEE Transactions on Computer-Aided Design of
  Integrated Circuits and Systems (TCAD)}}
  \bibinfo{volume}{27}(\bibinfo{number}{7}), pp. \bibinfo{pages}{1165--1178},
  \doi{10.1109/TCAD.2008.923410}.

\bibitemdeclare{inproceedings}{FHJLR07}
\bibitem{FHJLR07}
\bibinfo{author}{Ansgar \surnamestart Fehnker\surnameend},
  \bibinfo{author}{Ralf \surnamestart Huuck\surnameend},
  \bibinfo{author}{Patrick \surnamestart Jayet\surnameend},
  \bibinfo{author}{Michel \surnamestart Lussenburg\surnameend} \&
  \bibinfo{author}{Felix \surnamestart Rauch\surnameend}
  (\bibinfo{year}{2007}): \emph{\bibinfo{title}{Model Checking Software at
  Compile Time}}.
\newblock In: {\sl \bibinfo{booktitle}{Proceedings of the First Joint IEEE/IFIP
  Symposium on Theoretical Aspects of Software Engineering}},
  \bibinfo{series}{TASE '07}, \bibinfo{publisher}{IEEE Computer Society},
  \bibinfo{address}{Washington, DC, USA}, pp. \bibinfo{pages}{45--56},
  \doi{10.1109/TASE.2007.34}.

\bibitemdeclare{incollection}{Junker:2012}
\bibitem{Junker:2012}
\bibinfo{author}{Maximilian \surnamestart Junker\surnameend},
  \bibinfo{author}{Ralf \surnamestart Huuck\surnameend},
  \bibinfo{author}{Ansgar \surnamestart Fehnker\surnameend} \&
  \bibinfo{author}{Alexander \surnamestart Knapp\surnameend}
  (\bibinfo{year}{2012}): \emph{\bibinfo{title}{SMT-based False Positive
  Elimination in Static Program Analysis}}.
\newblock In \bibinfo{editor}{Toshiaki \surnamestart Aoki\surnameend} \&
  \bibinfo{editor}{Kenji \surnamestart Taguchi\surnameend}, editors: {\sl
  \bibinfo{booktitle}{14th International Conference on Formal Engineering
  Methods, Kyoto Japan}}, {\sl \bibinfo{series}{Lecture Notes in Computer
  Science}} \bibinfo{volume}{7635}, \bibinfo{publisher}{Springer Berlin
  Heidelberg}, pp. \bibinfo{pages}{316--331},
  \doi{10.1007/978-3-642-34281-3\_23}.

\bibitemdeclare{inproceedings}{queillesifakis82}
\bibitem{queillesifakis82}
\bibinfo{author}{Jean-Pierre \surnamestart Queille\surnameend} \&
  \bibinfo{author}{Joseph \surnamestart Sifakis\surnameend}
  (\bibinfo{year}{1982}): \emph{\bibinfo{title}{Specification and verification
  of concurrent systems in {CESAR}}}.
\newblock In: {\sl \bibinfo{booktitle}{Proc. Intl. Symposium on Programming,
  April 6--8}}, \bibinfo{publisher}{Springer Verlag}, pp.
  \bibinfo{pages}{337--350}, \doi{10.1007/3-540-11494-7\_22}.

\bibitemdeclare{book}{seacord2008cert}
\bibitem{seacord2008cert}
\bibinfo{author}{Robert~C. \surnamestart Seacord\surnameend}
  (\bibinfo{year}{2008}): \emph{\bibinfo{title}{The CERT C Secure Coding
  Standard}}, \bibinfo{edition}{1st} edition.
\newblock \bibinfo{publisher}{Addison-Wesley Professional}.

\end{thebibliography}

\end{document}